\documentclass[10pt,english,10pt]{revtex4}
\usepackage[T1]{fontenc}
\usepackage[latin1]{inputenc}
\usepackage{graphicx}

\makeatletter

\providecommand{\tabularnewline}{\\}

\makeatletter

\makeatletter

\usepackage{rotating}

\makeatletter

\makeatletter

\makeatletter

\usepackage{epsfig}

\makeatother

\makeatother

\makeatother

\makeatother
\def\lsim{\mathrel{\rlap{\lower4pt\hbox{\hskip1pt$\sim$}}
    \raise1pt\hbox{$<$}}}                

\makeatother

\usepackage{babel}
\makeatother
\begin{document}

\title{Measuring the dark side (with weak lensing)}

\author{Luca Amendola}

\address{INAF/Osservatorio Astronomico di Roma, Via Frascati 33, 00040 Monteporzio
Catone, Roma, Italy}

\author{Martin Kunz}

\address{Département de Physique Théorique, Université de Genève, 24 quai
Ernest Ansermet, CH--1211 Genève 4, Switzerland}

\author{Domenico Sapone}

\address{Département de Physique Théorique, Université de Genève, 24 quai
Ernest Ansermet, CH--1211 Genève 4, Switzerland}

\date{\today}

\begin{abstract}
We introduce a convenient parametrization of dark energy models that
is general enough to include several modified gravity models and generalized
forms of dark energy. In particular we take into account the linear
perturbation growth factor, the anisotropic stress and the modified
Poisson equation. We discuss the sensitivity of large scale weak lensing
surveys like the proposed DUNE satellite to these parameters. We find
that a large-scale weak-lensing tomographic survey is able to easily
distinguish the Dvali-Gabadadze-Porrati model from $\Lambda$CDM and
to determine the perturbation growth index to an absolute error of
$0.02-0.03$. 
\end{abstract}
\maketitle

\section{Introduction}

The observed late-time accelerated expansion (e.g. \cite{snls}) of
the Universe has opened a Pandora's box in cosmology. Although a great
number of models have emerged from the box, none of them has provided
a satisfactory explanation of the observations (\cite{tsuji} and
references therein).

In a landscape of hardly compelling theories, probably the most efficient
way to proceed is to exploit present and future data in search of
signatures of unexpected phenomena that may signal new physical effects.
In this way it is possible that we will be able to distinguish, say,
a cosmological constant from dynamical dark energy (DE) or the latter
from some form of modified gravity.

To this end, an important task is to provide observational groups
with simple measurable parameters that may be linked to interesting
physics. In this paper we will investigate the extent to which additional
parameters can be used to detect signatures of new cosmology in future
surveys, with particular emphasis on weak lensing (WL). These additional
parameters are connected to the growth of linear perturbations, to
the anisotropic stress (defined as the difference bewteen the Newtonian
potentials in longitudinal gauge) and to deviations from the Poisson
equation for matter. All these parameters find a simple motivation
in current models of modified gravity, from extradimensional ones
\cite{KuSa} to scalar-tensor theories; it is clear however that their
introduction is not limited to these cases and can in fact account
also for other phenomena, for instance clustering in the DE component.
We also discuss which of the parameters arise naturally in which context.
We then evaluate the sensitivity of WL experiments for two cases:
a phenomenological one, in which the parameterization is chosen mainly
on the grounds of simplicity and in analogy to some specific models;
and a more physically motivated case, namely the Dvali-Gabadadze-Porrati
(DGP) \cite{dgp} extra-dimensional model.

We focus on WL for two reasons: first, contrary to eg supernovae or
baryon-oscillation tests, WL makes use of both background and linear
perturbation dynamics, allowing to break degeneracies that arise at
the background level (this is particularly important for testing modified
gravity); second, several groups are planning or proposing large WL
experiments in the next decade (e.g.~\cite{DUNE,JDEM,SNAP}) that
will reach the sensitivity to test cosmology at unprecedented depth
and it is therefore important to optimize the science return of these
proposals. We will therefore produce Fisher matrix confidence regions
for the relevant parameters for surveys like those proposed by the
DUNE \cite{DUNE} or JDEM/SNAP \cite{JDEM,SNAP} collaborations.

The effect of the anisotropic stress has been discussed several times
in DE literature. Refs. \cite{acqua} and \cite{ScUzRi} discuss it
in the context of scalar-tensor theories and evaluated the effect
on the shear spectrum; \cite{acd} discusses its role for a coupled
Gauss-Bonnet theory; \cite{KuSa} showed that it is essential to mimic
the DGP growth-rate with a fluid dark energy model; finally, during
the final stages of this project another paper discussing the anisotropic
stress in dark energy models appeared \cite{CaCoMa}. Similarly, the
parametrization of the growth factor in DE models has been often investigated
in the past \cite{lahav,wang}.

\section{Defining the dark Side}

In this section we discuss our parametrisation of the dark sector.
We separate it into two parts, firstly the parametrisation of the
background quantities relevant for a perfectly homogeneous and isotropic
universe, and secondly the additional parameters which describe small
deviations from this idealised state, at the level of first order
perturbation theory.

We assume throughout that the spatial curvature vanishes, mostly for
simplicity. We note however that especially in the context of general
dark energy models curvature is not yet well constrained \cite{ClCoBa}.

\subsection{Parametrisation of the expansion history}

In order to characterise the evolution of the Universe at the background
level, we need to provide a parametrisation of the expansion history,
$H(z)$. To this end we write it as \begin{equation}
H(z)^{2}=H_{0}^{2}\left[\Omega_{m}(1+z)^{3}+(1-\Omega_{m})f(z)\right].\end{equation}
 We define $\Omega_{m}$ by the requirement that during matter domination,
i.e. at redshifts of a few hundred, the Universe expands according
to \begin{equation}
H(z\sim O(100))=H_{0}\sqrt{\Omega_{m}}(1+z)^{\frac{3}{2}}.\label{eq:Om}\end{equation}
 We will call this part the (dark) matter, and the remaining part
the dark energy. However, these are only effective quantities \cite{Kunz}.
It could be that we are dealing with a tracking dark energy that scales
like matter at high redshifts \cite{doran,DTV}. In this case, the
scaling part of its energy density would be counted as matter. It
is also possible that the observed acceleration of the expansion is
due to a modification of gravity. In this case there is no dark energy
at all and the $(1-\Omega_{m})f(z)$ part takes into account the gravity
terms. In all cases, the fundamental quantity is $H(z)$.

The function $f(z)$ describes the evolution of the energy density
in the (effective) dark energy. The most widely used parametrization
makes use of a Taylor series in $a$ \cite{chevpol}, \begin{equation}
w(a)=w_{0}+(1-a)w_{a}.\end{equation}
 where $a=1/(1+z)$ is the scale factor. For the above form of $w$
there is an analytic expression of $f$: \begin{equation}
f(z;w_{0},w_{a})=(1+z)^{3(1+w_{0}+w_{a})}\exp\left\{ -3w_{a}\frac{z}{1+z}\right\} \end{equation}
 However this is not necessarily a good fit to physical models or
to the data \cite{BaCoKu}; in the DGP case we will make use of the
full expression for $w(a)$.

\subsection{Parametrisation of the first order quantities}

We will limit ourselves to scalar perturbations. In Newtonian gauge
we can then write the line element defining the metric as \begin{equation}
ds^{2}=a(\tau)^{2}\left[-(1+2\psi)d\tau^{2}+(1-2\phi)dx^{2}\right]\end{equation}
 so that it is characterised by two perturbative quantities, the scalar
potentials $\psi$ and $\phi$.

The perturbations in the dark fluids are characterised for example
by their comoving density perturbations $\Delta=\delta+3HaV/k^{2}$
and their velocity perturbations $V$. Their evolution is sourced
by the potentials $\phi$ and $\psi$ and depends also on the pressure
perturbations $\delta p$ and the anisotropic stresses $\sigma$ of
the fluids. For the (dark) matter we set $\delta p=0$ (as in addition
$w_{m}=0$ this is a gauge-invariant statement) and $\sigma_{m}=0$.
In App. \ref{app:pert} we define our notation and review the perturbation
formalism in more detail.

In General Relativity the $\phi$ potential is given by the algebraic
relation \begin{equation}
-k^{2}\phi=4\pi Ga^{2}\sum_{i}\rho_{i}\Delta_{i},\label{eq:psi1}\end{equation}
 which is a generalisation of the Poisson equation of Newtonian gravity
(the factor $-k^{2}$ is the spatial Laplacian). We see that in general
all fluids with non-zero perturbations will contribute to it. Since
we have characterised the split into dark matter and dark energy at
the background level, we cannot demand in addition that $\Delta_{\mathrm{DE}}\ll\Delta{}_{m}$.
At the fluid level, the evolution of $\Delta_{\mathrm{DE}}$ is influenced
by a combination of the pressure perturbation and the anisotropic
stress of the dark energy. However, the pressure perturbation of the
dark energy is only very indirectly related to observables through
the Einstein equations. For this reason we rewrite Eq.~(\ref{eq:psi1})
as \begin{equation}
k^{2}\phi\equiv-4\pi Ga^{2}Q\rho_{m}\Delta_{m}.\label{eq:Q}\end{equation}
 where $G$ is the gravitational constant measured today in the solar
system. Here $Q(k,a)$ is a phenomenological quantity that, in general
relativity (GR), is due to the contributions of the non-matter fluids
(and in this case depends on their $\delta p$ and $\sigma$). But
it is more general, as it can describe a change of the gravitational
constant $G$ due to a modification of gravity (see DGP example below).
It could even be apparent: If there is non-clustering early quintessence
contributing to the expansion rate after last scattering then we added
its contribution to the total energy density during that period wrongly
to the dark matter, through the definition of $\Omega_{m}$. In this
case we will observe less clustering than expected, and we need to
be able to model this aspect. This is the role of $Q(k,a)$.

For the dark energy we need to admit an arbitrary anisotropic stress
$\sigma$, and we use it to parametrise $\psi$ as \begin{equation}
\psi\equiv[1+\eta(k,a)]\phi.\label{eq:eta}\end{equation}
 At present there is no sign for a non-vanishing anisotropic stress
beyond that generated by the free-streaming of photons and neutrinos.
However, it is expected to be non-zero in the case of topological
defects \cite{KuDu} or very generically for modified gravity models
\cite{KuSa}. In most cases, $\eta(a\rightarrow0)\rightarrow0$ in
order to recover the behaviour of the standard model, however in some
models like scalar-tensor theories this may not be the case.

If $Q\neq1$ or $\eta\neq0$, then we need to take into account the
modified growth of linear perturbations. Defining the logarithmic
derivative \begin{equation}
\frac{d\log\Delta_{m}}{d\log a}=m(a)\end{equation}
 it has been shown several times \cite{lahav,wang,amque,Linder,HuLi}
that for several DE models (including dynamical dark energy and DGP)
a good approximation can be obtained assuming\begin{equation}
m(a)=\Omega_{m}(a)^{\gamma}\end{equation}
 where $\gamma$ is a constant that depends on the specific model.

Although for an analysis of actual data it may be preferable to use
the parameters $\{ Q,\eta\}$, we concentrate in this paper on the
use of weak lensing to distinguish between different models. As we
will see in the following section, the quantities that enter the weak
lensing calculation are the growth index $\gamma$ as well as the
parameter combination \begin{equation}
\Sigma\equiv Q(1+\eta/2).\end{equation}
 Weak lensing will therefore most directly constrain these parameters,
so that we will use the set $\{\gamma,\Sigma\}$ for the constraints.
Of course we should really think of both as functions of $Q$ and
$\eta$. An additional benefit of using $\gamma$ is that as anticipated
it is relatively easy to parametrise since in most models it is reasonable
to just take $\gamma$ to be a constant (see section \ref{sec:gamma}).
For $\Sigma$ the situation is however not so clear-cut. First, a
simple constant $\Sigma$ would be completely degenerate with the
overall amplitude of the linear matter power spectrum, so would be
effectively unobservable per se with WL experiments. Then, even simple
models like scalar-tensor theories predict a complicate time-dependence
of $\Sigma$ so that it is not obvious which parametrization is more
useful (while for DGP it turns out that $\Sigma=1$ just as in GR,
see below). So lacking a better motivation we start with a very simple
possibility, namely that $\Sigma$ starts at early times as in GR
(i.e. $\Sigma=1$) and then deviates progressively more as time goes
by: \begin{equation}
\Sigma(a)=1+\Sigma_{0}a.\label{eq:s0}\end{equation}
 This choice is more general than it seems. If we assume that $\Sigma(a)$
is linear and equal to unity at some arbitrary $a_{1}$, then we have
$\Sigma(a)=1+\Sigma_{1}(a_{1}-a)=(1+\Sigma_{1}a_{1})[1-a\Sigma_{1}/(1+\Sigma_{1}a_{1})]$.
But since an overall factor can be absorbed into the spectrum normalization,
one has in fact $\Sigma(a)=1-\Sigma_{1}'a$ where $\Sigma_{1}'\equiv\Sigma_{1}/(1+\Sigma_{1}a_{1})$.
Therefore, from the error on $\Sigma_{0}$ in (\ref{eq:s0}) one can
derive easily the error on the slope $\Sigma_{1}$ at any point $a_{1}$.

As a second possibility we also investigate a piece-wise constant
function $\Sigma(a)$ with three different values $\Sigma_{1,2,3}$
in three redshift bins. Here again, due to the degeneracy with $\sigma_{8}$,
we fix $\Sigma_{1}=1$.

In this way we have parametrised the three geometric quantities, $H$,
$\phi$ and $\psi$, with the following parameters: $\Omega_{m}$,
$\Sigma_{0}$ (or $\Sigma_{2,3}$) and $\gamma$, plus those that
enter the effective dark energy equation of state. These represent
all scalar degrees of freedom to first order in cosmological perturbation
theory. The parameterisations themselves are clearly not general,
as we have replaced three functions by six or seven numbers, and as
we have not provided for a dependence on $k$ at all, but they work
well for $\Lambda$CDM, Quintessence models, DGP and to a more limited
extent for scalar-tensor models. We emphasize that this set is optimised
for weak lensing forecasts. For the analysis of multiple experiments,
one should use a more general parametrisation of $Q$ and $\eta$.
If the dark energy can be represented as a fluid, then $\{\delta p,\sigma\}$
would be another natural choice for the extra degrees of freedom,
with $\eta=\eta(\sigma)$ and $Q=Q(\delta p,\sigma)$. While phenomenological
quantities like $Q$, $\eta$, $\Sigma$ and $\gamma$ are very useful
for measuring the behaviour of the dark energy, we see them as a first
step to uncovering the physical degrees of freedom. Once determined,
they can guide us towards classes of theories -- for example $\eta=0$
would rule out many models where GR is modified.

\section{Observables}

\subsection{Constraining the expansion history}

In order to constrain the expansion history we can measure either
directly $H(z)$ or else one of the distance measures. The main tools
are

\begin{itemize}
\item \textbf{Luminosity distance}: Probed by type-Ia supernova explosions
(SN-Ia), this provides currently the main constraints on $H(z)$ at
low redshifts, $z\leq1.5$. 
\item \textbf{Angular diameter distance}: Measured through the tangential
component of the Baryon Acoustic Oscillations (BAO) either via their
imprint in the galaxy distribution \cite{sdss_bao} or the cosmic
microwave background radiation (CMB) \cite{wmap3}. One great advantage
of this method is the low level of systematic uncertainties. The CMB
provides one data point at $z\approx1100$, while the galaxy BAO probe
mostly the range $z\leq1.5$. Using Lyman-$\alpha$ observations this
could be extended to $z\approx3$. 
\item \textbf{Direct probes of H(z)}: This can either be done through the
radial component of the BAO in galaxies, or through the dipole of
the luminosity distance \cite{BoDuKu}. 
\end{itemize}

\subsection{Growth of matter perturbations\label{sec:gamma}}

In the standard $\Lambda$CDM model of cosmology, the dark matter
perturbations on sub-horizon scales grow linearly with the scale factor
$a$ during matter domination. During radiation domination they grow
logarithmically, and also at late times, when the dark energy starts
to dominate, their growth is slowed. The growth factor $g\equiv\Delta_{m}/a$
is therefore expected to be constant at early times (but after matter-radiation
equality) and to decrease at late times. In addition to this effect
which is due to the expansion rate of the universe, there is also
the possibility that fluctuations in the dark energy can change the
gravitational potentials and so affect the dark matter clustering.

In $\Lambda$CDM $g$ can be approximated very well through \begin{equation}
g(a)=\exp\left\{ \int_{0}^{a}d\ln a\left(\Omega_{m}(a)^{\gamma}-1\right)\right\} \label{eq:growth}\end{equation}
 where $\gamma\approx0.545$. There are two ways that the growth rate
can be changed with respect to $\Lambda$CDM: Firstly, a general $w$
of the dark energy will lead to a different expansion rate, and so
to a different Hubble drag. Secondly, if in the Poisson equation (\ref{eq:Q})
we have $Q\neq1$ then this will also affect the growth rate of the
dark matter, as will $\eta\neq0$. We therefore expect that $\gamma$
is a function of $w$, $\eta$ and $Q$.

For standard quintessence models we have $\eta=0$ and $Q\approx1$
on small scales as the scalar field does not cluster due to a sound
speed $c_{s}^{2}=1$. In this case $\gamma$ is only a weak function
of $w$ \cite{Linder}: \begin{equation}
\gamma(w)=0.55+0.05(1+w(z=1)).\label{eq:gwfit}\end{equation}
 A constant $\gamma$ turns out to be an excellent approximation also
for coupled dark energy models \cite{amque} and for modified gravity
models \cite{Linder}. A similar change in $\gamma$ can also be obtained
in models where the effective dark energy clusters \cite{KuSa}.

If we assume that $\psi$ is the dominant source of the dark matter
clustering, then we see that its value is modified by $(1+\eta)Q$
(while the $\dot{\phi}$ term sourcing perturbations in $\delta_{m}$
does not contain the contribution from the anisotropic stress). Following
the discussion in \cite{LiCa} we set \begin{equation}
A=\frac{(1+\eta)Q-1}{1-\Omega_{m}(a)}\end{equation}
 and derive the asymptotic growth index as \begin{equation}
\gamma_{\infty}=\frac{3(1-w_{\infty}-A(Q,\eta))}{5-6w_{\infty}}.\end{equation}
 In section \ref{sec:dgp} on the DGP model we show how these relations
can be inverted to determine $\phi$ and $\psi$ .

Although these relations are useful to gain physical insight and for
a rough idea of what to expect, for an actual data analysis one would
specify $Q$ and $\eta$ and then integrate the perturbation equations.

\subsection{Weak lensing \label{sec:WL}}

Usually it is taken for granted that matter concentrations deflect
light. However, the light does not feel the presence of matter, it
feels the gravitational field generated by matter. In this paper we
consider a scenario where the gravitational field has been modified,
so that we have to be somewhat careful when deriving the lensing equations.
Following \cite{ScUzRi} we exploit the fact that it is the lensing
potential $\Phi=\phi+\psi$ which describes the deviation of light
rays in our scenario. In $\Lambda$CDM we can use the Poisson equation
(\ref{eq:psi1}) to replace the lensing potential with the dark matter
perturbations (since the cosmological constant has no perturbations),
\begin{equation}
k^{2}\Phi=2\frac{3H_{0}^{2}\Omega_{m}}{2a}\Delta_{m}.\end{equation}
 However, in general this is {\em not} the complete contribution
to the lensing potential. For our parametrisation we find instead
that \begin{equation}
k^{2}\Phi=Q(2+\eta)\frac{3H_{0}^{2}\Omega_{m}}{2a}\Delta_{m}=2\Sigma\frac{3H_{0}^{2}\Omega_{m}}{2a}\Delta_{m}.\end{equation}
 Correspondingly, the total lensing effect is obtained by multiplying
the usual equations by a factor $\Sigma=(1+\eta/2)Q$. We also notice
that a modification of the growth rate appears \emph{twice}, once
in the different behaviour of $\Delta_{m}$, and a second time in
the non-trivial factor $\Sigma$. It is clear that we need to take
both into account, using only one of the two would be inconsistent.

An additional complication arises because some of the scales relevant
for weak lensing are in the (mildly) non-linear regime of clustering.
It is therefore necessary to map the linear power spectrum of $\Phi$
into a non-linear one. This is difficult even for pure dark matter,
but in this case fitting formulas exist, and numerical tests with
$N$-body programs have been performed. For non-standard dark energy
models or modified gravity the mapping is not known, and it probably
depends on the details of the model. Although this may be used in
the long run as an additional test, for now we are left with the question
as to how to perform this mapping. In this paper we have decided to
assume that $\Sigma\Delta_{m}$ represents the effective clustering
amplitude, and we use this expression as input to compute the non-linear
power spectrum.

One side effect of this prescription is that as anticipated any constant
pre-factor of $\Delta_{m}$ becomes degenerate with $\sigma_{8}$
(which is also a constant) and cannot be measured with weak lensing.
In practice this means that if we use $\{\gamma,\Sigma\}$ as the
fundamental parameters, then a constant contribution to $\Sigma$
cannot be measured. On the other hand if we use $\{ Q,\eta\}$ as
fundamental parameters then even a constant $\eta$ affects $\gamma$
and the degeneracy with $\sigma_{8}$ is broken through this link.

\subsection{Other probes}

We have already mentioned one CMB observable, namely the peak position,
which is sensitive mostly to the expansion rate of the universe and
provides effectively a standard ruler. On large scales the CMB angular
power spectrum is dominated by the integrated Sachs-Wolfe (ISW) effect,
which is proportional to $\dot{\phi}+\dot{\psi}=\dot{\Phi}$. The
ISW effect therefore probes a similar quantity as weak lensing, but
is sensitive to its time evolution. Additionally, it is mostly relevant
at low $\ell$, which means that it is strongly affected by cosmic
variance. This limits its statistical power in a fundamental way.

{} From the perturbation equation for the matter velocity perturbation,
Eq.~(\ref{eq:v_m}) we see that $V$ is only sensitive to $\psi$.
Additionally, the peculiar velocities of galaxies are supposed to
be a very good tracer of the dark matter velocity field \cite{pecvel}.
The peculiar velocities are therefore a direct probe of $\psi$ alone.
This makes it in principle an excellent complement of weak lensing
(which measures $\phi+\psi$) and of the growth rate (measuring mostly
$\phi$ through the Poisson equation) but of course measuring reliably
peculiar velocities to cosmological distances is still prohibitive.

The perturbations in the metric also affect distance measurements
like e.g.~the luminosity distance \cite{BoDuGa,HuGr}. The fluctuations
in the luminosity distance on small angular scales are a measure of
both the peculiar motion of the supernovae as well as the lensing
by intervening matter perturbations. As very large supernova data
sets are expected in the future, this may turn into a promising additional
probe of $\phi$ and $\psi$.

\section{Dark energy models}

Let us now review some of the models among those presented in the
literature that are amenable to our parametrization.

\subsection{Lambda-CDM}

In $\Lambda$CDM the dark energy contribution to the energy momentum
tensor is of the form $T_{\mu}^{\nu}=\Lambda\delta_{\mu}^{\nu}$.
{} From this we see immediately that $p=-\rho$ so that we have a
constant $w=-1$. Additionally there are no perturbations in a cosmological
constant, which is compatible with the perturbation equations (\ref{eq:delta})
and (\ref{eq:v}) as they decouple from the gravitational potentials
for $w=-1$. For this reason $Q=1$. The absence of off-diagonal terms
in the energy momentum tensor also shows that $\eta=0$ so that $\Sigma=1$.
The growth-rate of the matter perturbations is only affected by the
accelerated expansion of the universe, with $\gamma\approx6/11$.

\subsection{Quintessence}

In quintessence models the dark energy is represented by a scalar
field, so that $w$ can now vary as a function of time, subject to
the condition $w\geq-1$ if $\rho_{Q}>0$. At the level of first order
perturbations, quintessence is exactly equivalent to a fluid with
$c_{s}^{2}=1$ and no anisotropic stress ($\eta=0$). For such a fluid
one often relaxes the condition on $w$, so that $w<-1$ becomes allowed
(see e.g. \cite{KuSa06} and papers cited therein for more details).
The high sound speed suppresses clustering on sub-horizon scales so
that $|Q(k\gg H_{0})-1|\ll1$. It is on these scales that weak lensing
and galaxy surveys measure the matter power spectrum, so that $\Sigma\approx1$,
and the growth-rate is only slightly changed through the different
expansion rate. From (\ref{eq:gwfit}) we see that typically one has
$\gamma=0.54\div0.57$ for the range of $w$ still allowed by the
data.

On very large scales clustering can be non-negligible, which affects
for example the low $\ell$ part of the CMB spectrum through the ISW
effect \cite{WeLe}. For a DGP-like equation of state, we find $Q(a=1,k=H_{0})\lsim1.1$
by integrating the perturbation equations numerically. In general
we expect that $Q-1$ remains small below the sound horizon, while
it can be non-negligible on larger scales. Thus, if we find that $Q\approx1$
on small scales, but $Q>1$ for scales larger than some critical scale,
then this may be a hint for the existence of a sound horizon of the
dark energy. Additionally, one important class of quintessence models
exhibits a period of {}``tracking'' at high redshift, during which
the energy density in the dark energy stays constant relative to the
one in the dark matter \cite{doran,DTV}. If the dark energy makes
up an important fraction of the total energy density during that period,
$\Omega_{m}$ as defined through Eq.~(\ref{eq:Om}) will be too large.
In that case $Q<1$ as the quintessence clusters less than the dark
matter on small scales.

\subsection{A generic dark energy model}

A generic non-standard model with either a dark energy very different
from quintessence or in which gravity is non-Einsteinian will presumably
introduce modifications in the Poisson equation, the equation for
dark matter growth and/or the anisotropic stress. Since these quantities
are not independent, a generic modified gravity model should allow
for at least an additional parameter for $\Sigma$ and one for $\gamma$.
As we anticipated we assume $\gamma$ constant and then either $\Sigma=1+\Sigma_{0}a$
or a piece-wise constant $\Sigma$ in three redshift bins. Let us
denote these two {}``generic dark energy'' models as GDE1 and GDE2. 

Since these parametrisations contain $\Lambda$CDM in their parameter
space, which is the phenomenologically most successful model today,
they are useful to characterise the sensitivity of experiments to
non-standard dark energy models.

\subsection{DGP\label{sec:dgp}}

\begin{figure}[ht]
 \center\epsfig{figure=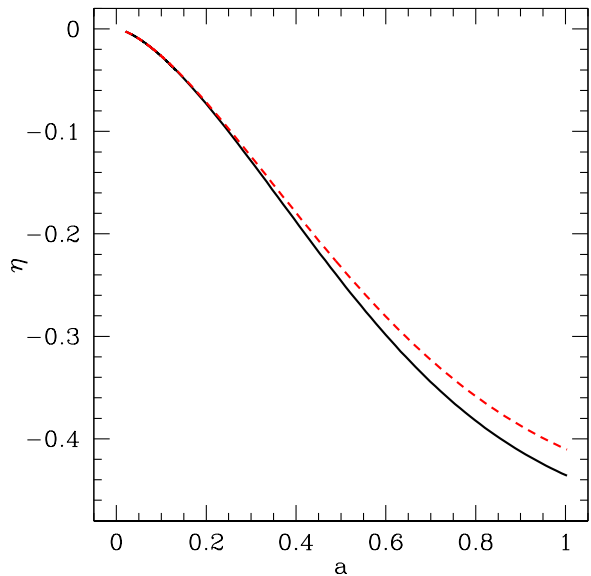,width=3.2in}

\caption{The $\eta$ parameter of the anisotropic stress as a function of
$a$ for $\Omega_{m}=0.3$. The black solid line shows the actual
value while the red dashed curve shows the recovered $\eta$ using
the fitting formula for $\gamma$.}

\label{fig:eta_dgp} 
\end{figure}

An alternative approach to the late-time accelerated expansion of
the universe modifies the geometry side of the Einstein equations,
rather than the energy content. A well-known model of this kind is
the Dvali-Gabadadze-Porrati model \cite{dgp}. This model is based
on five-dimensional gravity, with matter and an additional four-dimensional
Einstein-Hilbert action on a brane. This then modifies the evolution
of the Universe, with one solution asymptotically approaching a de
Sitter universe. Assuming spatial flatness, the Hubble parameter for
this solution is given by \cite{MaMa} \begin{equation}
H^{2}=\frac{H}{r_{c}}+\frac{8\pi G}{3}\rho_{m}.\end{equation}
 We can solve this quadratic equation for $H$ to find \begin{equation}
H=\frac{1}{2r_{c}}+\sqrt{\frac{1}{4r_{c}^{2}}+\frac{8\pi G}{3}\rho_{m}}.\end{equation}
 Since the matter stays on the brane, its conservation equation is
four-dimensional so that $\rho_{m}\propto a^{-3}$. Considering an
effective dark energy component with $\rho_{\mathrm{eff}}\equiv3H/(8\pi Gr_{c})$
which leads to the DGP expansion history, one can use the conservation
equation $\dot{\rho}_{\mathrm{eff}}+3H(1+w_{\mathrm{eff}})\rho_{\mathrm{eff}}=0$
to define an effective equation of state with \cite{MaMa} \begin{equation}
w_{\mathrm{eff}}(a)=\frac{\Omega_{m}-1-\sqrt{(1-\Omega_{m})^{2}+4\Omega_{m}/a^{3}}}{2\sqrt{(1-\Omega_{m})^{2}+4\Omega_{m}/a^{3}}}.\label{eq:w_dgp}\end{equation}
 For $\Omega_{m}=0.3$ we find $w_{0}\approx-0.77$ and $w_{a}\approx0.3$.
For $\eta$ we turn to \cite{LuScSt,KoMa} which have computed the
perturbations on small scales and found \begin{eqnarray}
k^{2}\phi & = & -4\pi G\, a^{2}\left(1-\frac{1}{3\beta}\right)\rho_{m}\Delta_{m}\label{eq:dgp_phi}\\
k^{2}\psi & = & -4\pi G\, a^{2}\left(1+\frac{1}{3\beta}\right)\rho_{m}\Delta_{m}\label{eq:dgp_psi}\end{eqnarray}
 where the parameter $\beta$ is defined as: \begin{equation}
\beta=1-2Hr_{c}\left(1+\frac{\dot{H}}{3H^{2}}\right)=1-\frac{2(Hr_{c})^{2}}{2Hr_{c}-1}=1+2Hr_{c}w_{\mathrm{DE}}\end{equation}
 With the parametrisation Eq.~(\ref{eq:eta}) we find that \begin{equation}
\eta=\frac{2}{3\beta-1}=\frac{\Omega_{m}(a)^{2}-1}{2+\Omega_{m}(a)^{2}}.\label{eq:eta_dgp}\end{equation}
 which means that its value today is $\eta(a=1)\approx-0.44$ for
$\Omega_{m}=0.3$. We plot its evolution as a function of $a$ in
Fig.~\ref{fig:eta_dgp}. We see that it vanishes at high redshift,
when the modifications of gravity are negligible. %

\begin{figure}[ht]
 \center\epsfig{figure=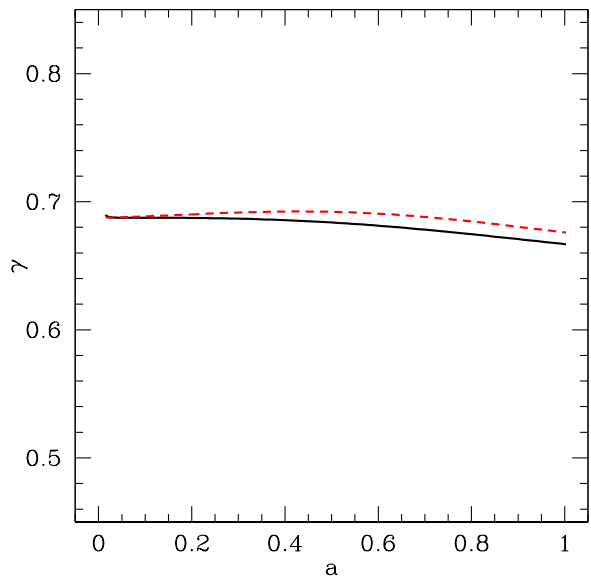,width=3.2in}

\caption{The growth parameter $\gamma$ of DGP, comparison between the fitting
formula (red dashed curve) and the numerical result (black solid line)
for $\Omega_{m}=0.3$.}

\label{fig:gamma_dgp} 
\end{figure}

For the growth function we turn to \cite{LiCa}. They find \begin{equation}
\gamma=\frac{7+5\Omega_{m}(a)+7\Omega_{m}^{2}(a)+3\Omega_{m}^{3}(a)}{[1+\Omega_{m}^{2}(a)][11+5\Omega_{m}(a)]}.\end{equation}
 Fig.~\ref{fig:gamma_dgp} compares this formula with the numerical
result, and we see that it works very well. On average in the range
$z\in(0,3)$ we can use $\gamma\approx0.68$. %

\begin{figure}[ht]
 \center\epsfig{figure=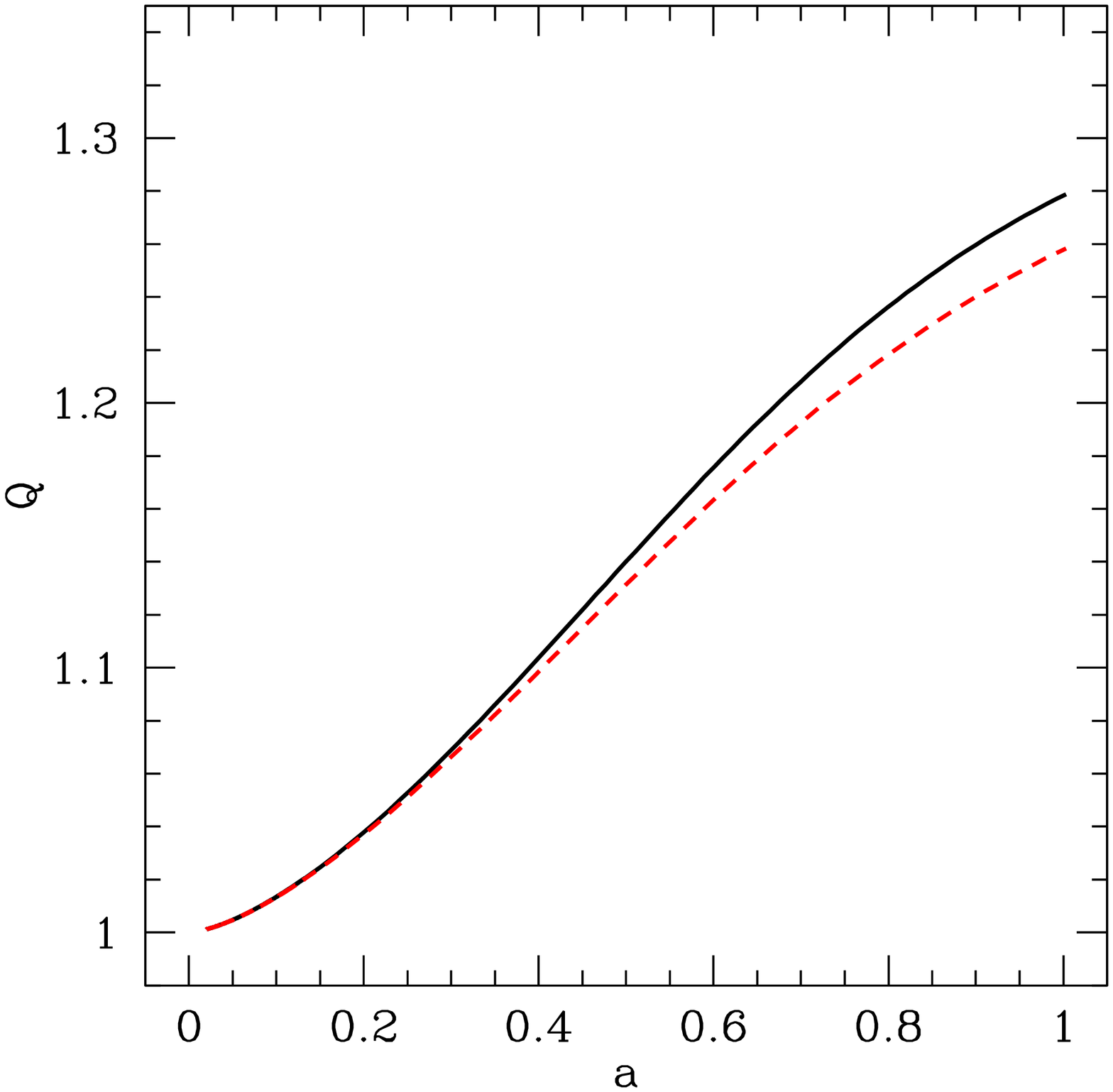,width=3.2in}

\caption{The $Q$ parameter of DGP, $\Omega_{m}=0.3$. The black solid curve
shows the exact value while the red dashed line is the result recovered
with the fitting formula for $\gamma$.}

\label{fig:q_dgp} 
\end{figure}

$Q$ is just the non-trivial pre-factor in Eq.~(\ref{eq:dgp_phi}),
\begin{equation}
Q=1-\frac{1}{3\beta},\end{equation}
 it is plotted in Figure \ref{fig:q_dgp}. We find that for DGP $\Sigma=1$
so that the deviation of light rays due to a given mass is the same
as in GR. The weak lensing results therefore depend only on $\gamma$
which modifies the growth of $\Delta_{m}$, but not on $Q$ and $\eta$
separately. In this argument we also used that the gravitational constant
$G_{\mathrm{cav}}$ measured by a Cavendish type experiment is just
the bare constant $G$ as the force-law modifications of DGP are screened
for $\Delta_{m}\gg1$ \cite{LuScSt}. This is in contrast to the situation
for scalar-tensor theories.

It is maybe instructive to illustrate with a short {}``Gedanken''
experiment how we could recover $\phi$ and $\psi$ from data if our
universe was described by a DGP-like model. We assume that $w(a)$
has been measured by e.g. supernovae and/or BAO. Suppose now that
looking at the growth-history of the matter power spectrum (available
potentially as a by-product from a BAO survey or from a dedicated
galaxy survey) we notice that $\gamma$ is outside the range allowed
for quintessence-like models. We therefore have to assume that the
dark energy is non-standard. We now need to invert the computation
of $\gamma$ to get the link with $\phi$, through the definition
of $Q(a)$, \begin{equation}
k^{2}\phi=-4\pi Ga^{2}Q(a)\rho_{m}\Delta_{m}.\label{eq:defq}\end{equation}
 Using the relations in section \ref{sec:gamma} we find \begin{equation}
(1+\eta(a))Q(a)-1=\left(1-\Omega_{m}(a)\right)\left(1-w(a)-\frac{5-6w(a)}{3}\gamma(a)\right).\label{eq:etaQ}\end{equation}
 Lensing on the other hand gives \begin{equation}
k^{2}(\phi+\psi)=k^{2}\phi(2+\eta)=-(2\Sigma)4\pi Ga^{2}\rho_{m}\Delta_{m}.\end{equation}
 For DGP the lensing signal is precisely the one expected naively
for $\Delta_{m}$, so that \begin{equation}
\Sigma=1\Leftrightarrow\eta=\frac{2}{Q}-2\end{equation}
 We find that $(1+\eta)Q=2-Q$, allowing us to recover $Q$ and $\eta$
separately. They are shown as the dashed curves in Figs.~\ref{fig:eta_dgp}
and \ref{fig:q_dgp}. Once we know $\eta$, $\gamma$ and $Q$ we
can compute $\phi$ and $\psi$.

We find that the accuracy of the fitting formula for $\gamma(Q,\eta)$
is quite good, and certainly sufficient for current experiments, and
for distinguishing DGP from Quintessence at the perturbation level.
We will see later that next-generation weak lensing experiment can
reach a precision where the differences are important. At that point
one may need to numerically integrate the perturbation equations to
compute $\gamma$.

\subsection{$\Lambda$DGP}

We find it instructive to introduce a simple variant of DGP that we
denote as $\Lambda$DGP, namely a DGP model which includes a cosmological
constant. In this way we can interpolate between DGP proper ($\Omega_{\Lambda}=0$)
and $\Lambda$CDM ($\Omega_{\Lambda}=1-\Omega_{m}$, ie $\Omega_{r_{c}}=0$).
Notice that we are still taking the self-accelerating branch of DGP,
different from e.g. \cite{LuSt}. In flat space, the Hubble parameter
is then given by \begin{equation}
H^{2}=\frac{H}{r_{c}}+\frac{8\pi G}{3}\left(\rho_{m}+\rho_{\Lambda}\right)\end{equation}
 with the self-accelerating solution now being \begin{equation}
H=\frac{1}{2r_{c}}+\sqrt{\frac{1}{4r_{c}^{2}}+\frac{8\pi G}{3}\left(\rho_{m}+\rho_{\Lambda}\right)}=H_{0}\left[\sqrt{\Omega_{r_{c}}}+\sqrt{\Omega_{r_{c}}+\Omega_{\Lambda}+\Omega_{m}a^{-3}}\right].\label{hubble}\end{equation}
 In flat space we additionally have that $\Omega_{r_{c}}=\frac{1}{4}\left(1-\Omega_{m}-\Omega_{\Lambda}\right)^{2}$.
We also notice that we can define an overall effective dark energy
fluid through \begin{equation}
\rho_{\mathrm{eff}}=\rho_{\Lambda}+\frac{3H}{8\pi Gr_{c}}.\end{equation}
 from which we can derive an effective equation of state. Concerning
the perturbations, it was shown in \cite{LuSt} that the DGP force
laws (\ref{eq:dgp_phi}) and (\ref{eq:dgp_psi}) are not changed through
the addition of a brane cosmological constant if we write $\beta$
as \begin{equation}
\beta=1-2r_{c}H\left(1+\frac{\dot{H}}{3H^{2}}\right)\end{equation}
 which depends on the value of $\Lambda$. For our key quantities
we find \begin{eqnarray}
w(a) & = & -\frac{1-\Omega_{m}\left(a\right)+\Omega_{\Lambda}\left(a\right)}{\left[1-\Omega_{m}\left(a\right)\right]\left[1+\Omega_{m}\left(a\right)+\Omega_{\Lambda}\left(a\right)\right]}\\
\beta\left(a\right) & = & -\frac{\left[1+\Omega_{m}\left(a\right)+\Omega_{\Lambda}\left(a\right)\right]^{2}-2\Omega_{m}\left(a\right)}{\left[1-\Omega_{m}\left(a\right)-\Omega_{\Lambda}\left(a\right)\right]\left[1+\Omega_{m}\left(a\right)+\Omega_{\Lambda}\left(a\right)\right]}\\
\eta\left(a\right) & = & \frac{1-\left[\Omega_{m}\left(a\right)+\Omega_{\Lambda}\left(a\right)\right]^{2}}{3\Omega_{m}\left(a\right)-\left[1+\Omega_{m}\left(a\right)+\Omega_{\Lambda}\left(a\right)\right]\left[2+\Omega_{m}\left(a\right)+\Omega_{\Lambda}\left(a\right)\right]}\\
Q(a) & = & \frac{2}{3}\frac{1+\Omega_{\Lambda}\left(a\right)-2\Omega_{m}\left(a\right)+\left(1+\Omega_{m}\left(a\right)+\Omega_{\Lambda}\left(a\right)\right)^{2}}{\left(1+\Omega_{m}\left(a\right)+\Omega_{\Lambda}\left(a\right)\right)^{2}-2\Omega_{m}\left(a\right)}\\
\Sigma(a) & = & 1\end{eqnarray}
 These reduce to the ones of DGP and $\Lambda$CDM in the respective
limits. The growth factor can be approximated by \begin{equation}
\gamma\left(a\right)=\frac{\left[\left(1+\Omega_{m}+\Omega_{\Lambda}\right)^{2}-2\Omega_{m}\right]\left[7-4\Omega_{m}+5\Omega_{\Lambda}-3\Omega_{m}^{2}-3\Omega_{m}\Omega_{\Lambda}\right]+2\Omega_{m}\left(1-\Omega_{m}-\Omega_{\Lambda}\right)}{\left[\left(1+\Omega_{m}+\Omega_{\Lambda}\right)^{2}-2\Omega_{m}\right]\left[\left(11+5\Omega_{m}\right)\left(1-\Omega_{m}\right)+\left(11-5\Omega_{m}\right)\Omega_{\Lambda}\right]}\end{equation}
 where for ease of notation we suppressed the explicit dependence
of the $\Omega$s on $a$ .

\subsection{Scalar-tensor theories}

For completeness we also give a brief overview of the relevant quantities
in scalar-tensor theories (see eg. \cite{acqua,ScUzRi}). For a model
characterized by the Lagrangian \begin{equation}
L=F(\varphi)R-\varphi_{;\mu}\varphi^{\,;u}-2V(\varphi)+16\pi G^{*}L_{matter}\end{equation}
 (where $F(\varphi)$ is the coupling function, that we assume to
be normalized to unity today and $G^{*}$ is the bare gravitational
constant) the relation between the metric potentials is \begin{equation}
\psi=\phi-\frac{F'}{F}\delta\varphi\end{equation}
 where $F'=dF/d\varphi$. It turns out that in the linear sub-horizon
limit the functions $\psi,\phi,\delta\varphi$ obey three Poisson-like
equations: \begin{eqnarray}
k^{2}\phi & = & -4\pi\frac{G^{*}}{F}a^{2}\rho_{m}\Delta_{m}\frac{2(F+F'^{2})}{2F+3F'^{2}}\\
k^{2}\psi & = & -4\pi\frac{G^{*}}{F}a^{2}\rho_{m}\Delta_{m}\frac{2(F+2F'^{2})}{2F+3F'^{2}}\label{eq:psipoisson}\\
k^{2}\delta\varphi & = & 4\pi\frac{G^{*}}{F}a^{2}\rho_{m}\Delta_{m}\frac{2FF'}{2F+3F'^{2}}\end{eqnarray}
Then from (\ref{eq:defq}) we derive\begin{equation}
Q=\frac{G^{*}}{FG_{cav,0}}\frac{2(F+F^{\prime2})}{2F+3F^{\prime2}}\end{equation}
 where $G_{cav,0}$ is the presently measured value of the gravitational
constant in a Cavendish-like experiment. If the equation (\ref{eq:psipoisson})
can be assumed to hold in the highly non-linear laboratory environment
then one would define \begin{equation}
G_{cav,0}=\frac{G^{*}}{F_{0}}\frac{2(F_{0}+2F_{0}^{\prime2})}{2F_{0}+3F_{0}^{\prime2}}\end{equation}
 Moreover, we obtain the anisotropic stress\begin{equation}
\eta\equiv\frac{\psi-\phi}{\phi}=\frac{F^{\prime2}}{F+F^{\prime2}}\end{equation}
 Finally, we derive \begin{equation}
\Sigma\equiv Q\left(1+\frac{\eta}{2}\right)=\frac{G^{*}}{FG_{cav,0}}\end{equation}
 (notice that our result differs from \cite{ScUzRi}). It is clear
then that depending on $F$ our simple phenomenological parametrization
may be acceptable or fail completely. Moreover we find that the usual
growth fit (\ref{eq:growth}) is not a very good approximation since
during the matter era the growth is faster than in a $\Lambda$CDM
model. The analysis of specific examples of scalar-tensor models is
left to future work.

\begin{table}
\begin{tabular}{|c|c|c|c|c|}
\hline 
Model&
growth index&
$\Sigma(a)$&
new param.&
fid. values\tabularnewline
\hline
\hline 
GDE1&
$\gamma=\mbox{const}$&
$\Sigma=1+\Sigma_{0}a$&
$\gamma,\Sigma_{0}$&
$(0.55,0)$\tabularnewline
\hline 
GDE2&
$\gamma=\mbox{const}$&
$\Sigma(\mbox{i-th }z\mbox{-bin})=\Sigma_{i}$&
$\gamma,\Sigma_{2},\Sigma_{3}$&
$(0.55,1,1)$\tabularnewline
\hline 
DGP&
$\gamma=\mbox{const}$&
$\Sigma=1$&
$\gamma$&
$0.68$\tabularnewline
\hline 
$\Lambda$DGP&
$\gamma(a)$&
$\Sigma=1$&
$\Omega_{\Lambda}$&
$0$ or $0.7$\tabularnewline
\hline
\end{tabular}

\caption{The DE models considered in this paper.}
\end{table}

\section{Forecasts for weak lensing large-scale surveys}

We finally are in position to derive the sensitivity of typical next-generation
tomographic weak lensing surveys to the non-standard parameters introduced
above, expanding over recent papers like Refs. \cite{jain} and \cite{taylor}.
In particular, we study a survey patterned according to the specifications
in Ref. \cite{amref}, which dealt with the standard model. In Appendix
B we give the full convergence power spectrum as a function of $\eta$
and $Q$ and the relevant Fisher matrix equations.

Let us then consider a survey characterized by the sky fraction $f_{\mathrm{sky}}$,
the mean redshift $z_{\mathrm{mean}}\approx1.412z_{0}$ and the number
sources per arcmin$^{2}$, $d$. When not otherwise specified we assume
$z_{\mathrm{mean}}=0.9$ and $d=35$ as our benchmark survey: these
values are well within the range considered for the DUNE satellite
proposal. The derived errors scale clearly as $f_{\mathrm{sky}}^{1/2}$
so it is easy to rescale our results to different sky fractions. We
assume that the photo-$z$ error obeys a normal distribution with
variance $\sigma_{z}=0.05$. We choose to bin the distribution out
to $z=3$ into five equal-galaxy-number bins (or three for the model
with a piece-wise constant $\Sigma(z)$). For the linear matter power
spectrum we adopt the fit by Eisenstein \& Hu \cite{ehu} (with no
massive neutrinos and also neglecting any change of the shape of the
spectrum for small deviations around $w=-1$). For the non-linear
correction we use the halo model by Smith et al. \cite{smith}. We
consider the range $10<\ell<20000$ since we find that both smaller
$\ell$ and larger $\ell$'s do not contribute significantly.

We begin the discussion with the generic dark energy models GDE1 and
GDE2. The parameter set (with the fiducial values inside square brackets)
is therefore\begin{equation}
p_{\alpha}=\{\omega_{m}\equiv\Omega_{m}h^{2}[0.147],\omega_{b}\equiv\Omega_{b}h^{2}[0.02205],n_{s}[1],\Omega_{m}[0.3],w_{0}[-0.95],w_{a}[0],\gamma[0.55],\sigma_{8}[0.8]\}\end{equation}
 while for $\Sigma$ we assume as fiducial values either $\Sigma_{0}=0$
(GDE1) or $\Sigma_{1,2,3}=1$ (GDE2).

\begin{figure}
\includegraphics[clip,scale=0.8]{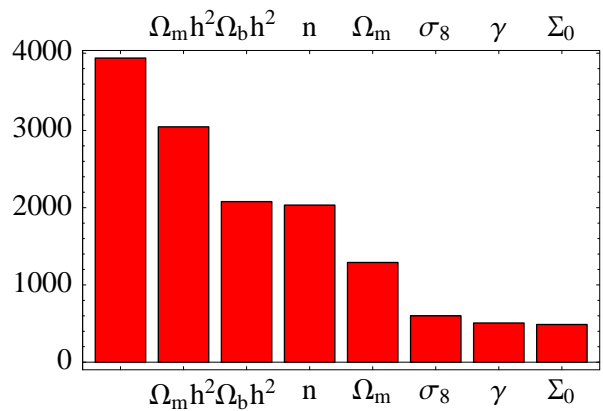}

\caption{\label{fig:fom}FOM for $w_{0},w_{p}$vs. marginalized parameters
of the model GDE1.}
\end{figure}

First we study how the estimate of $w_{0},w_{p}$ (projection of $w_{0},w_{a}$
on the pivot point $a_{p}$ defined as the epoch at which the errors
decorrelate) is affected by fixing the other parameters. In Fig. \ref{fig:fom}
we show the FOM defined as $1/[\sigma(w_{0})\cdot\sigma(w_{p})]$
first when all the parameters are fixed to their fiducial value (first
bar) and then successively marginalizing over the parameter indicated
in the label and over all those on the left (eg the fourth column
represents the marginalization over $\omega_{m},\omega_{b},n_{s}$).
This shows that the WL method would benefit most from complementary
experiments that determine $\Omega_{m},\Omega_{m}h^{2}$. On the other
hand, there is not much loss in marginalizing over the two non-standard
parameters $\Sigma_{0},\gamma$.

\begin{figure}
\includegraphics[clip,scale=0.8]{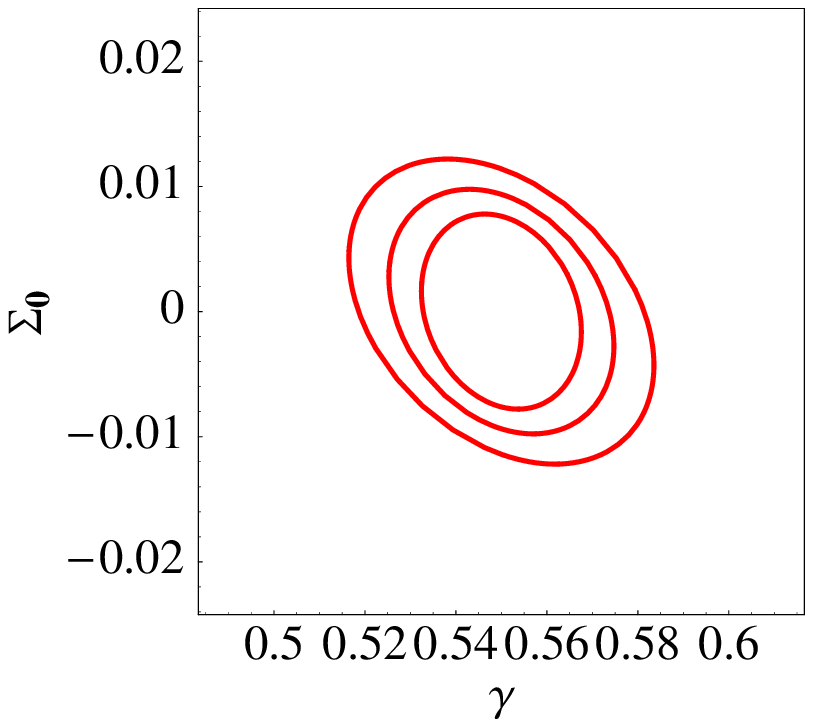}

\caption{\label{fig:cl}Confidence regions at 68\% for the benchmark survey
$z_{mean}=0.9,d=35$ (outer contour) and for $d=50,75$ (inner contours).}
\end{figure}

\begin{figure}
\includegraphics[clip,scale=0.8]{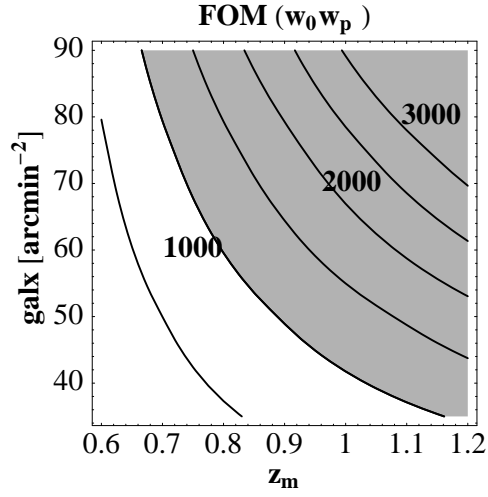}

\caption{\label{fig:FOM2}FOM ($w_{0},w_{p}$) vs. $z_{mean}$ and $d$ (galaxies
per arcmin$^{2}$) in GDE1. The grey area represents a convenient
target. }
\end{figure}

In Fig. \ref{fig:cl} we show the confidence regions for $\Sigma_{0},\gamma$
. Errors of the order or 0.1 for $\Sigma_{0}$ and 0.3 for $\gamma$
are reachable already with the benchmark survey. In Fig. \ref{fig:FOM2}
we show the FOM ($w_{0},w_{p}$) varying the depth $z_{mean}\approx1.412z_{0}$
and the density $d$ of sources per arcmin$^{2}$ (full marginalization).
If we set as a convenient target a FOM equal to 1000 (for instance,
an error of 0.01 for $w_{0}$ and 0.1 for $w_{p}$) then we see that
our benchmark survey remains a little below the target (we obtain
$\sigma(w_{0})=0.018$ and $\sigma(w_{p})=0.088$), which would require
at least $d=50$ or a deeper survey.

\begin{figure}
\includegraphics[clip,scale=0.8]{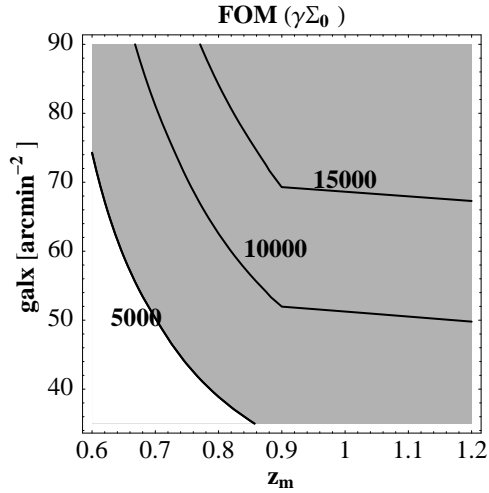}

\caption{\label{fig:FOM3}FOM ($\gamma,\Sigma_{0}$) vs. $z_{mean}$ and $d$
(galaxies per arcmin$^{2}$) in GDE1. The target is within the grey
area. }
\end{figure}

In Fig. \ref{fig:FOM3} we show the FOM ($\gamma,\Sigma_{0}$) again
varying $z_{mean},d$ (full marginalization). Here we set as target
a FOM of 5000, obtained for instance with an error 0.02 on $\gamma$
and $0.01$ on $\Sigma_{0}$; it turns out that the target can be
reached with the benchmark survey.

\begin{figure}
\includegraphics[clip,scale=0.8]{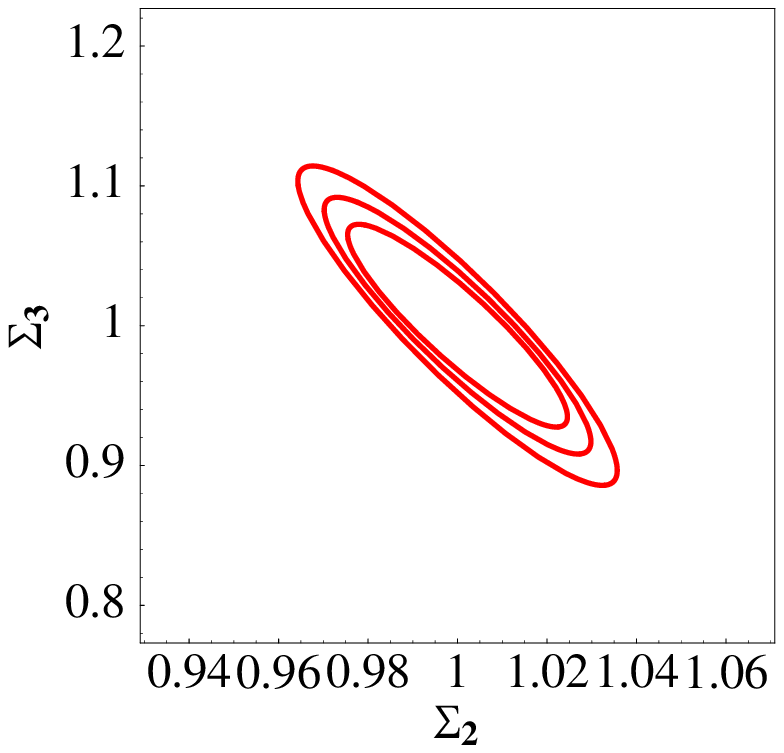}

\caption{\label{fig:sigma23}Confidence regions at 68\% for the benchmark
survey $z_{mean}=0.9,d=35$ (outer contour) and for $d=50,75$ (inner
contours) in the model GDE2.}
\end{figure}

For the model GDE2 we divide the survey into three equal-galaxy-numbers
$z$-bins and choose a $\Sigma(z_{i})=\Sigma_{i}$ piece-wise constant
in the three bins. Fixing $\Sigma_{1}=1$, we are left with two free
parameters $\Sigma_{2,3}$. In Fig. \ref{fig:sigma23} we show the
confidence regions; we see that WL surveys could set stringent limits
on the deviation of $\Sigma$ from the GR fiducial value.

\begin{figure}
\includegraphics[clip,scale=0.8]{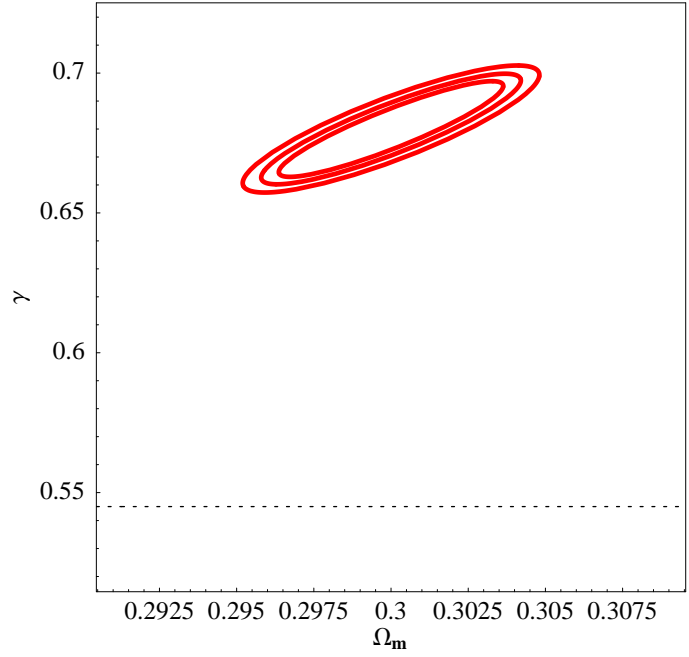}

\caption{\label{fig:dgp}Confidence regions at 68\% for the benchmark survey
$z_{mean}=0.9,d=35$ (outer contour) and for $d=50,75$ (inner contours)
for DGP. The dotted line represents the $\Lambda$CDM value.}
\end{figure}

We can now focus our attention to the DGP model. As anticipated, in
order to investigate the ability of WL studies to distinguish the
DGP model from $\Lambda$CDM, we consider two cases. First, we assume
a standard DGP model with $w(z)$ given by Eq.~(\ref{eq:w_dgp}).
In this case the model also determines the function $\gamma(z)$.
For $\Omega_{m}\approx0.3$ one has an almost constant $\gamma$ in
the range $z\in(0,3)$ with an average value $\gamma\approx0.68$.
Instead of using the full equation for $\gamma(z)$ we prefer to leave
$\gamma$ as a free constant parameter in order to compare directly
with a standard gravity DE model with the same $w(z)$ and the standard
value $\gamma\approx0.55$. In Fig. \ref{fig:dgp} we show the confidence
regions around the DGP fiducial model; our benchmark surveys seems
well capable of differentiating DGP from $\Lambda$CDM.

\begin{figure}
\includegraphics[clip,scale=0.8]{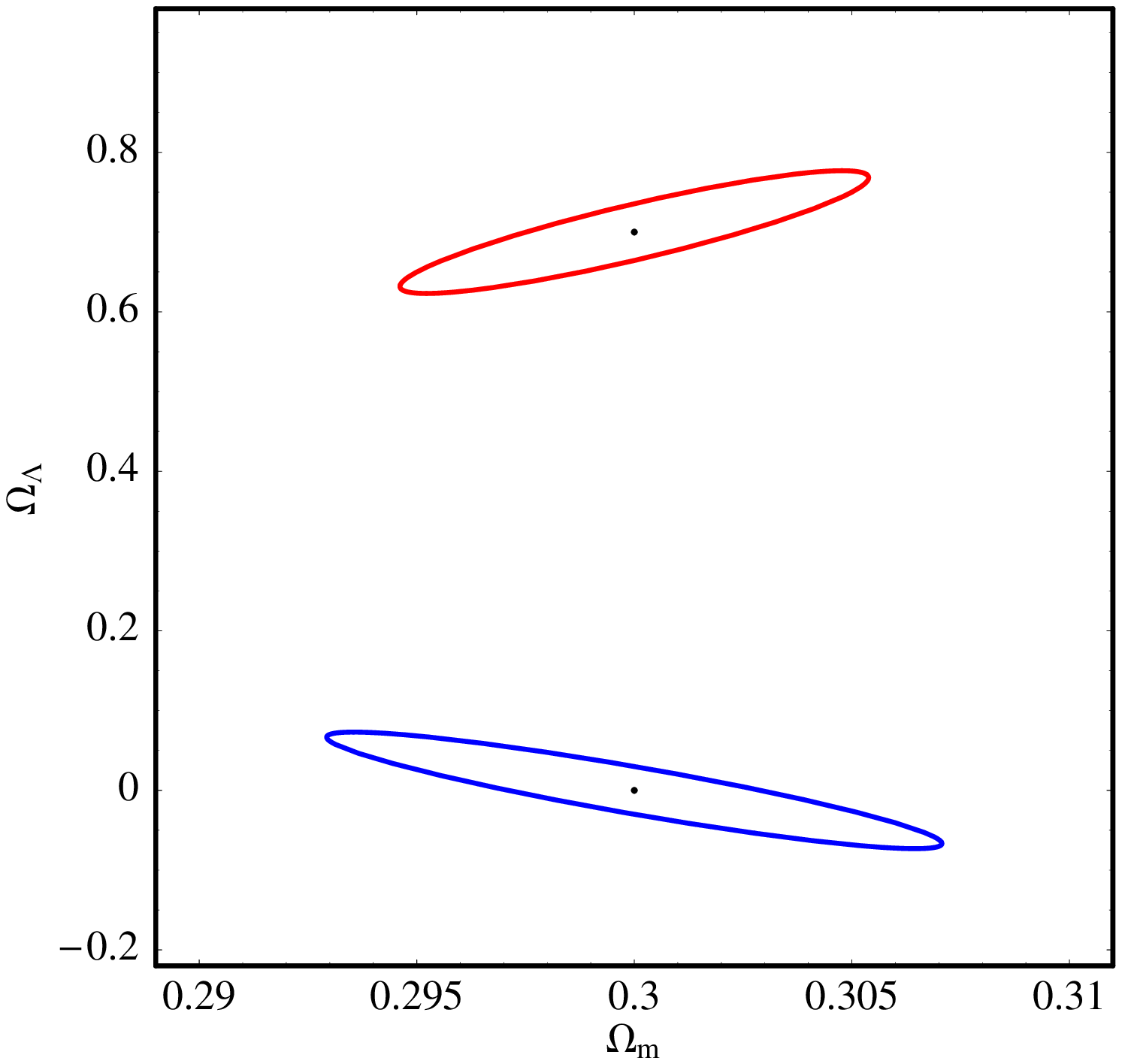}

\caption{\label{fig:LDGP}Confidence regions at 68\% for the benchmark survey
$z_{mean}=0.9,d=35$ for the model $\Lambda$DGP assuming as fiducial
value $\Omega_{\Lambda}=0$ (below) and $\Omega_{\Lambda}=1-\Omega_{m}$
(above).}
\end{figure}

\begin{figure}
\includegraphics[clip,scale=0.68]{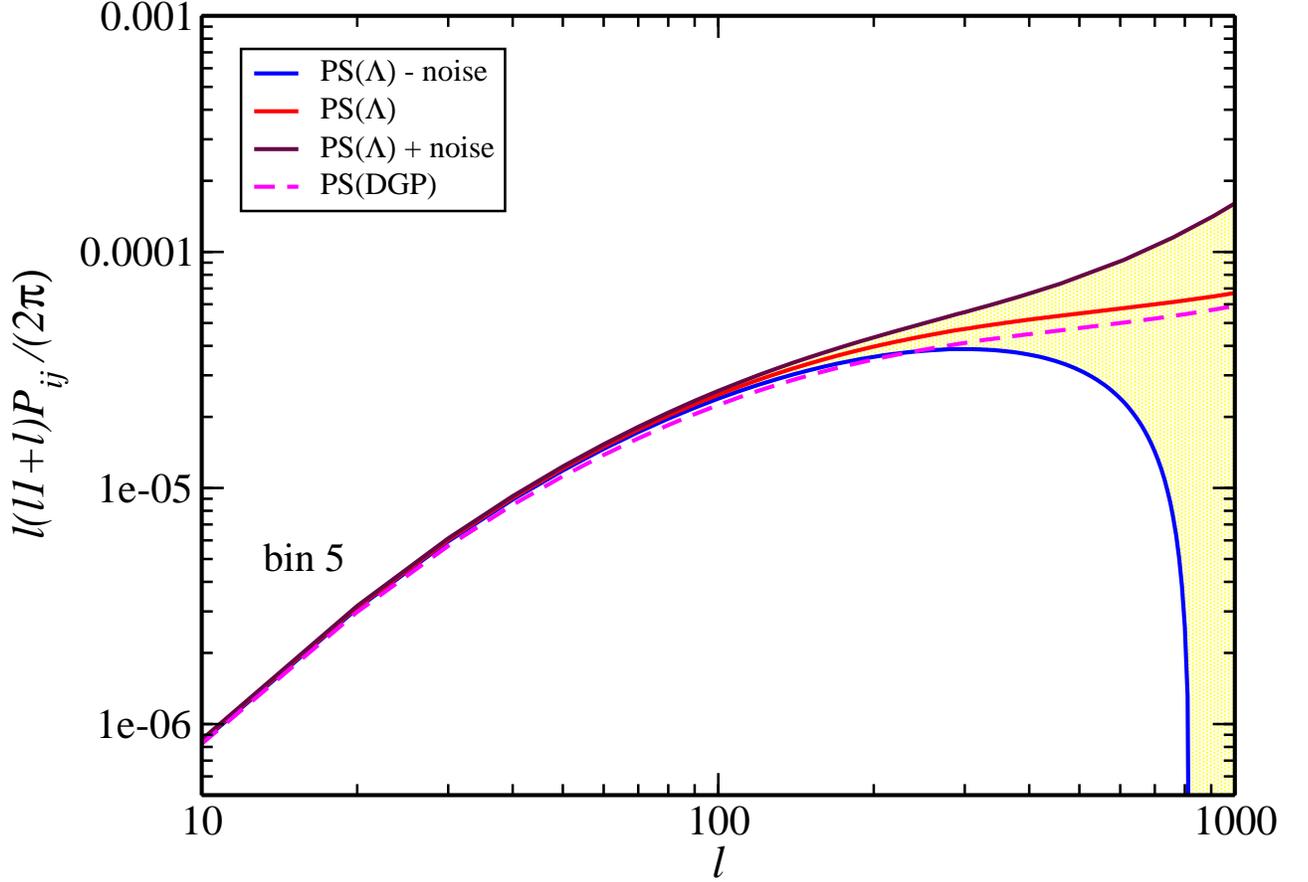}

\caption{\label{fig:CnlPS} The non linear power spectrum for $\Lambda$CDM
and DGP model for one single bin (bin 5). The red (central) solid
line and the magenta dashed line show the convergence power spectrum
for the $\Lambda$CDM and the DGP model respectively. The shaded area
(delimited by solid blue and brown lines) shows the noise errors on
the $\Lambda$CDM convergence power spectrum.}
\end{figure}

Then, we consider the $\Lambda$DGP model, in which the {}``matter''
content is in fact matter plus a cosmological constant, so that in
the limit of $\Omega_{\Lambda}=0$ one recovers DGP, while when $\Omega_{\Lambda}=1-\Omega_{m}$
one falls back into pure $\Lambda$CDM. In Fig.~\ref{fig:LDGP} we
see again that our benchmark survey will be able to distinguish between
the two extreme cases with a very high confidence. In Fig. \ref{fig:CnlPS}
we display the weak lensing spectrum for $\Lambda$CDM in the 5th
$z$-bin with the noise due to the intrinsic ellipticity and for comparison
the DGP spectrum. We see that the DGP spectrum is well outside the
noise at low $\ell$'s.

\section{Conclusions}

In this paper we propose a general parametrisation of both dark energy
and modified gravity models up to the linear perturbations. Apart
from parametrising the Hubble parameter $H(z)$ with an effective
equation of state, we use the growth index $\gamma$ and the effective
modification of the lensing potential $\Sigma$. We discuss the relation
of these quantities to the anisotropic stress (parametrised through
$\eta$ or $\sigma$) and the modification of the Poisson equation
(given in terms of a parameter $Q$). We then show how these parameters
appear in different experimental setups, concentrating specifically
on the case of weak lensing. We also give explicit expressions for
the parameters for a range of models like $\Lambda$CDM, Quintessence,
the DGP model and scalar-tensor theories. We identify a few signatures
that could point to specific theories: The detection of a significant
anisotropic stress would favour modified-gravity like theories, while
strong upper limits on $|\eta|$ could rule out many such models.
A significant deviation from $Q=1$ on large scales only could point
to a finite sound speed of the dark energy. The hope is that we can
eventually use such clues to understand the physical nature of the
phenomenon underlying the accelerated expansion of the Universe.

We use our parametrisation to provide forecasts for weak lensing satellite
experiments (having in mind a setup similar to DUNE) on how well they
will be able to constrain dark energy and modified gravity models.
We find that a DUNE-like survey will be able to constrain the growth
index with an error that varies from 0.015 to 0.036 depending on the
model (see Table \ref{tab:t2}). This is sufficient to rule out a
model like DGP at more than 7 standard deviations based on the perturbations.
Table \ref{tab:t2} also shows that the parameter $\Sigma$ can be
strongly constrained. This demonstrates that weak lensing will evolve
in the next decade into a very powerful probe of the dark energy phenomenon,
with the potential to deliver insight into the physics behind the
accelerated expansion of the universe through constraints on the dark
energy perturbations and on modified gravity.

%
\begin{table}
\begin{tabular}{|c|c|}
\hline 
Model&
constraints\tabularnewline
\hline
\hline 
GDE1&
$\sigma(\gamma)=0.022;\quad\sigma(\Sigma_{a})=0.008$\tabularnewline
\hline 
GDE2&
$\sigma(\gamma)=0.036;\quad\sigma(\Sigma_{2})=0.024;\quad\sigma(\Sigma_{3})=0.076$\tabularnewline
\hline 
DGP&
$\sigma(\gamma)=0.015$\tabularnewline
\hline 
$\Lambda$DGP&
$\sigma(\Omega_{\Lambda})_{\Lambda CDM}=0.051;\quad\sigma(\Omega_{\Lambda})_{DGP}=0.049$\tabularnewline
\hline
\end{tabular}

\caption{Constraints on the parameters in terms of the standard deviation
$\sigma$ (benchmark survey, fully marginalized). }

\label{tab:t2} 
\end{table}

\subsubsection*{Acknowledgments}

MK and DS acknowledge funding from the Swiss NSF. We thank Viviana
Acquaviva, Carlo Baccigalupi, Ruth Durrer, Valeria Pettorino and Carlo
Schimd for interesting discussions.

\appendix

\section{Perturbation equations\label{app:pert}}

In this appendix we recapitulate the basics of of first order perturbation
theory, both to collect the most important equations in one place
and to define our notation. We limit ourselves to scalar perturbations,
so that we can use the Newtonian gauge (also known as longitudinal
gauge), following \cite{MaBe}. The metric perturbations are then
defined by two scalar potentials $\psi$ and $\phi$, \begin{equation}
ds^{2}=a^{2}\left[-\left(1+2\psi\right)d\tau^{2}+\left(1-2\phi\right)dx_{i}dx^{i}\right]\label{pert_newton_ds}\end{equation}

The total energy momentum-tensor for a perfect fluid is given by:
\begin{equation}
T^{\mu\nu}=\left(\rho+p\right)u^{\mu}u^{\nu}+p~g^{\mu\nu}\label{EMT}\end{equation}
 where $\rho$ and $p$ are the density and the pressure of the fluid
respectively and the $u^{\mu}$ is the four-velocity. We parametrise
the averaged pressure via an equation of state parameter $w$, \begin{equation}
p=w\rho.\label{EOS}\end{equation}
 The perturbed energy-momentum tensor in Newtonian gauge can be written
as \begin{eqnarray}
T_{0}^{0} & = & -\left(\rho+\delta\rho\right)\label{00_EM}\\
ikT_{0}^{i} & = & -ikT_{i}^{0}=\left(1+w\right)\rho\theta\label{0i_EM}\\
T_{j}^{i} & = & \left(p+\delta p\right)\delta_{j}^{i}+\Pi_{j}^{i}\label{ij_EM}\end{eqnarray}
 where $\theta$ can be thought of as the divergence of a velocity
field, and $\Pi_{j}^{i}=T_{j}^{i}-\delta_{j}^{i}T_{k}^{k}/3$, the
traceless component of the space-space part of the energy momentum
tensor, represents an anisotropic stress. The scalar part of the anisotropic
stress $\sigma$ is related to $\Pi_{j}^{i}$ through \begin{equation}
\left(\rho+p\right)\sigma=-\left(\hat{k}_{i}\hat{k}_{j}-\frac{1}{3}\delta_{j}^{i}\right)\Pi_{j}^{i}\end{equation}
 We additionally introduce a different velocity variable $V=(1+w)\theta$
which is better behaved at $w=-1$ \cite{KuSa06}. In these variables
the perturbation equations for the fluids are \begin{eqnarray}
\delta_{i}' & = & 3(1+w_{i})\phi'-\frac{V_{i}}{Ha^{2}}-\frac{3}{a}\left(\frac{\delta p_{i}}{\rho_{i}}-w_{i}\delta_{i}\right)\label{eq:delta}\\
V_{i}' & = & -(1-3w_{i})\frac{V_{i}}{a}+\frac{k^{2}}{Ha^{2}}\frac{\delta p_{i}}{\rho_{i}}+(1+w_{i})\frac{k^{2}}{Ha^{2}}\psi-\frac{k^{2}}{Ha^{2}}\sigma_{i}.\label{eq:v}\end{eqnarray}
 where the prime denotes the derivative with respect to the scale
factor $a$. The fluid perturbations are all linked by their coupling
to the gravitational potentials, \begin{eqnarray}
 &  & k^{2}\phi=-4\pi Ga^{2}\sum_{i}\left[\rho_{i}\delta_{i}+3\frac{\dot{a}}{a}\frac{\rho_{i}V_{i}}{k^{2}}\right]\label{pert-delta/Eins}\\
 &  & k^{2}\left(\phi-\psi\right)=12\pi Ga^{2}\sum_{i}\left(\rho_{i}+p_{i}\right)\sigma_{i}\label{pert-sigma-Eins}\end{eqnarray}
 where the first equation is a combination of the $0-0$ and $0-i$
Einstein equations of \cite{MaBe}. It is often convenient to replace
the density contrast $\delta$ by the comoving density perturbation,
\begin{equation}
\Delta=\delta+3\frac{\dot{a}}{a}\frac{V}{k^{2}}\label{com-delta}\end{equation}
 and the pressure perturbation $\delta p$ is often parametrised with
the rest-frame sound speed $c_{s}^{2}$, \begin{equation}
\delta p=c_{s}^{2}\delta\rho+3aH\left(c_{s}^{2}-c_{a}^{2}\right)\rho\frac{V}{k^{2}}\end{equation}
 where $c_{a}^{2}=\dot{p}/\dot{\rho}$ is the adiabatic sound speed.

As an example, collisionless cold dark matter has zero pressure ($w_{m}=0$),
vanishing sound speed $c_{s,m}^{2}=c_{a,m}^{2}=0$ and no anisotropic
stress $\sigma_{m}=0$. The perturbation equation for the matter fluid
then become: \begin{eqnarray}
\delta' & = & 3\phi'-\frac{V}{Ha^{2}}\label{eq:delta_m}\\
V' & = & -\frac{V}{a}+\frac{k^{2}}{Ha^{2}}\psi.\label{eq:v_m}\end{eqnarray}

\section{The lensing Fisher matrix}

Here we discuss how the lensing Fisher matrix is modified in the general
case. Let us briefly recall the main equations for weak lensing studies.
The convergence weak lensing power spectrum can be written as \cite{hujain}
\begin{equation}
P_{ij}(\ell)=H_{0}^{3}\int_{0}^{\infty}\frac{dz}{E(z)}W_{i}(z)W_{j}(z)P_{nl}\left[P_{l}\left(\frac{H_{0}\ell}{r(z)},z\right)\right]\end{equation}
 where $P_{nl}[P_{l}(k,z)]$ is the non-linear matter power spectrum
at redshift $z$ obtained correcting the linear power spectrum $P_{l}(k,z)$.
In flat space we have:\begin{eqnarray}
W_{i} & = & \frac{3}{2}\Omega_{m}F_{i}(z)(1+z)\\
\nonumber \\F_{i}(z) & = & \int_{Z_{i}}dz_{s}\frac{n_{i}(z_{s})r(z,z_{s})}{r(0,z_{s})}\\
n_{i}(z) & = & D_{i}(z)/\int_{0}^{\infty}D_{i}(z')dz'\\
H(z) & = & H_{0}E(z)\\
r(z,z_{s}) & = & \int_{z}^{z_{s}}\frac{dz'}{E(z')}\end{eqnarray}
 where $D_{i}(z)$ is the radial distribution function of galaxies
in the $i$-th $z$-bin. We assume an overall radial distribution
\begin{equation}
D(z)=z^{2}\exp[-(z/z_{0})^{1.5}]\end{equation}
 The distributions $D_{i}$ are obtained by binning the overall distribution
and convolving with the photo-$z$ distribution function.

The Fisher matrix for weak lensing is given by\begin{equation}
F_{\alpha\beta}=f_{sky}\sum_{\ell}\frac{(2\ell+1)\Delta\ell}{2}\partial(P_{ij})_{,\alpha}C_{jk}^{-1}\partial(P_{km})_{,\beta}C_{mi}^{-1}\end{equation}
 where the cosmological parameters are $p_{\alpha}$ and partial derivatives
represent $\partial/\partial p_{\alpha}$, and \begin{equation}
C_{jk}=P_{jk}+\delta_{jk}\left\langle \gamma_{int}^{2}\right\rangle n_{j}^{-1}\end{equation}
 where $\gamma_{int}$ is the rms intrinsic shear (we assume $\left\langle \gamma_{int}^{2}\right\rangle ^{1/2}=0.22$
\cite{amref}) and \begin{equation}
n_{j}=3600d(\frac{180}{\pi})^{2}\hat{n}_{j}\end{equation}
 is the number of galaxies per steradians belonging to the $i$-th
bin, $d$ being the number of galaxies per square arcminute and $\hat{n}_{i}$
the fraction of sources belonging to the $i$-th bin.

As we have seen in the main text, we can parametrize a large number
of modified gravity models by the linear growth factor and by the
combined effect of the modified Poisson equation and the anisotropic
stress (the function $\Sigma$). So we have that the convergence spectrum
can be written as \begin{equation}
P_{ij}(\ell)=H_{0}^{3}\int_{0}^{\infty}\frac{dz}{E(z)}W_{i}(z)W_{j}(z)P_{nl}\left[Q^{2}(1+\frac{\eta}{2})^{2}P_{l}\left(\frac{H_{0}\ell}{r(z)},z\right)\right]\end{equation}
 where the functions $\eta$ and $Q$ will in general depend on $k$
(and therefore on $\ell$) and $z$. Notice moreover that the matter
power spectrum $P_{nl}$ depends on the linear growth function which
itself is a function of the background expansion $H(z)$ and of the
functions $Q(k,z)$ and $\eta(k,z)$.

\end{document}